\documentclass[12pt,superscriptaddress,aps,prd,preprint,showpacs]{revtex4}

\usepackage[dvips]{graphicx}
\usepackage{amsfonts}
\usepackage{slashed}
\usepackage[latin1]{inputenc}
\usepackage{amsmath}
\usepackage{hyperref}

\begin{document}

\title{Derivative four-fermion model, effective action and bumblebee generation}

\author{R. Ara\'ujo}
\affiliation{Instituto de F\'\i sica, Universidade Federal de Alagoas,\\ 57072-900, Macei\'o, Alagoas, Brazil}
\email{raline.araujo,tmariz@fis.ufal.br}

\author{T. Mariz}
\affiliation{Instituto de F\'\i sica, Universidade Federal de Alagoas,\\ 57072-900, Macei\'o, Alagoas, Brazil}
\email{raline.araujo,tmariz@fis.ufal.br}

\author{J. R. Nascimento}
\affiliation{Departamento de F\'{\i}sica, Universidade Federal da Para\'{\i}ba,\\
 Caixa Postal 5008, 58051-970, Jo\~ao Pessoa, Para\'{\i}ba, Brazil}
\email{jroberto,petrov@fisica.ufpb.br}

\author{A. Yu. Petrov}
\affiliation{Departamento de F\'{\i}sica, Universidade Federal da Para\'{\i}ba,\\
 Caixa Postal 5008, 58051-970, Jo\~ao Pessoa, Para\'{\i}ba, Brazil}
\email{jroberto,petrov@fisica.ufpb.br}

\begin{abstract}

In this paper, we study the one-loop effective potential of a derivative four-fermion model. As a result, an exact bumblebee-like potential is radiatively generated. Afterwards, we generalize our study for a finite temperature case and explicitly demonstrate the possibility of phase transitions allowing for the restoration of the Lorentz symmetry. We also investigate the low-energy effective action, from which we obtain the usual kinetic term and the corresponding bumblebee potential.

\end{abstract}


\maketitle

\section{Introduction}

The idea of dynamical symmetry breaking was first introduced by Bjorken in \cite{Bjorken:1963vg} (for an updated review, see \cite{Bjorken:2001pe}). The study focused on a four-fermion model with a self-interacting term $j_\mu j^\mu$, where $j^\mu=\bar\psi\gamma^\mu\psi$. As a continuation of this concept, various issues related to dynamical symmetry breaking have been studied in many further studies, among which it worth to mention \cite{Bialynicki-Birula:1963usy,Guralnik:1964zz,Phillips:1966zzc,Eguchi:1976iz,Amati:1981xt,Andrianov:1998ay,Kraus:2002sa,Jenkins:2003hw}, emphasizing especially the role of the outstanding contribution of Eguchi \cite{Eguchi:1976iz}. We note that a similar methodology has been adopted earlier to generate the effective dynamics of the scalar and gauge fields, including the curved space-time case \cite{Inagaki:1997kz} and possible supersymmetric extension \cite{Buchbinder:1997ta}, as well as higher-derivative couplings \cite{Elizalde:1998sv}. Also, this approach has been applied to generate the effective dynamics of the antisymmetric tensor field \cite{Leblanc:1993gx}. 

The concept of dynamical symmetry breaking has found its natural application within the context of Lorentz-violating (LV) theories. Indeed, the spontaneous Lorentz symmetry breaking, first,  already in the first papers where the concept of Lorentz symmetry breaking (LSB) itself was formulated \cite{Kostelecky:1989jp,Kostelecky:1989jw}, was assumed to be a very natural mechanism for arising constant LV vectors (tensors) allowing to define new terms within the LV Standard Model Extension (LV SME) defined in \cite{ColKost1,ColKost2} (further this idea was also corroborated in \cite{Andrianov:1998ay}), second, is the most natural manner to introduce LSB in gravity \cite{KosGra} where the explicit LSB faces various difficulties (see \cite{KosLi} for a discussion). We note that within the LV framework, the LV terms in many papers are shown to arise as radiative corrections in some theories where the fermions are coupled to the electromagnetic field with some LV constant vectors or tensors, with further these vectors (or tensors) being integrated out (see \cite{ourLV} and references therein). The first example of such an approach has already been presented in \cite{ColKost2}. In other words, one can say that the LV terms are generated as quantum corrections in models where the measurable fields, namely gauge and scalar ones, are coupled to some primordial spinor field, which is afterwards integrated out.

The dynamical LSB within the bumblebee framework has been studied on the base of a four-fermion interaction model with the chiral current $j_5^\mu=\bar\psi\gamma^\mu\gamma_5\psi$, in the one-loop approximation \cite{Gomes:2007mq} (we note that there exists a certain analogy between the axial vector coupling, which, in particular, is employed in our paper, and the chiral derivative coupling considered in \cite{Schwinger}, while, namely, the axial vector coupling is necessary to generate the bumblebee Lagrangian). The study showed that radiative corrections could generate a bumblebee-like potential defined in \cite{Kostelecky:1989jp,Kostelecky:1989jw,Kostelecky:2000mm,Altschul:2005mu}, i.e., a possibility of generating spontaneous Lorentz symmetry breaking through quantum corrections was demonstrated. However, the bumblebee potential vanishes when the 't Hooft-Veltman prescription is applied \cite{tHooft:1972tcz}. On the other hand, if we consider massless fermions, an explicit generation of the complete one-loop bumblebee potential   also can be performed \cite{Assuncao:2017tnz}.

More recently, in \cite{Assuncao:2019azw}, a four-fermion model with the derivative current $j^{\mu\nu}=\bar\psi\gamma^\mu i\partial^\nu\psi$ was analyzed, generating a tensor bumblebee potential in the nonperturbative approach. This finding has encouraged further research into other four-fermion models that may produce exact bumblebee potentials. Therefore, the objective of this work is to investigate a particular derivative self-interacting model that can generate an exact vector bumblebee potential.

The paper is organized as follows. In Section \ref{potential}, we show that a vector bumblebee potential may be induced through radiative corrections in the nonperturbative approach from a derivative four-fermion model. In Section \ref{action}, we obtain the kinetic term and the bumblebee potential in the perturbative approach. Finally, in Section \ref{summary}, we provide a summary of our results.

\section{Effective potential}\label{potential}

The four-fermion model that we are interested in is described by the Lagrangian
\begin{equation}
{\cal L}_0 = \bar\psi(i\slashed{\partial} - m)\psi -\frac{G}{2}\big(\bar\psi\textstyle{1\over2}\{i\hspace{-0.03cm}\overset{\leftrightarrow}{\slashed{\partial}},\gamma^\mu\}\gamma_5\psi\big)^2.
\end{equation}
In this model, the derivative current is expressed as $j_5^\mu = \bar\psi\textstyle{1\over2}\{i\hspace{-0.03cm}\overset{\leftrightarrow}{\slashed{\partial}},\gamma^\mu\}\gamma_5\psi$, where $\overset{\leftrightarrow}{\partial}_{\!\mu} = \partial_\mu - \overset{\leftarrow}{\partial}_{\!\mu}$ represents the left-right derivative. We can simplify this expression by considering $\textstyle{1\over2}\{i\hspace{-0.03cm}\overset{\leftrightarrow}{\slashed{\partial}},\gamma^\mu\} = \overset{\leftrightarrow}{i\partial_\mu}$, which leads to a new form of the Lagrangian: 
\begin{equation}\label{L0}
{\cal L}_0 = \bar\psi(i\slashed{\partial} - m)\psi -\frac{G}{2}\big(\bar\psi\overset{\leftrightarrow}{i\partial_\mu}\gamma_5\psi\big)^2.
\end{equation}

We note that, unlike \cite{Assuncao:2017tnz}, we consider massive fermions, which, in principle, can make the exact spinor propagator and, consequently, all studies more complicated compared to the massless case. Another difference in our model lies in the fact that our current, and hence the four-fermion vertex, involves derivatives, and even the dimension of our coupling constant differs from that in \cite{Assuncao:2017tnz}. Therefore, our model is essentially distinct from the last one.

As usual, we can introduce an auxiliary field to eliminate the self-interaction term of Eq.~(\ref{L0}). This involves rewriting the Lagrangian as 
\begin{eqnarray}
{\cal L} &=& {\cal L}_0 + \frac{g^2}{2} \big(B_\mu-\frac{e}{g^2}\bar\psi\overset{\leftrightarrow}{i\partial_\mu}\gamma_5\psi\big)^2 \nonumber \\
&=& \frac{g^2}{2}B_\mu B^\mu + \bar\psi(i\slashed{\partial}-m)\psi - eB^\mu(\bar\psi\overset{\leftrightarrow}{i\partial_\mu}\gamma_5\psi),
\end{eqnarray}
where we have introduced the new vector field $B_{\mu}$ and defined $G=\frac{e^2}{g^2}$. By applying the left-right derivative, we can also express the Lagrangian as
\begin{eqnarray}
{\cal L} &=& \frac{g^2}{2}B_\mu B^\mu+\bar\psi(i\slashed{\partial}-m-ie(\partial_\mu B^\mu)\gamma_5-2ieB^\mu\partial_\mu\gamma_5)\psi.
\end{eqnarray}
We note that, in principle, one could expect the arising of anomalies in our model due to the presence of the $\gamma_5$ matrix, which in principle can allow to generate terms proportional to $\epsilon^{\mu\nu\lambda\rho}B_{\mu\nu}B_{\lambda\rho}$ on the base of the mechanism analogous to that one described in \cite{Adler:1969gk,Bell:1969ts}. However, in our paper, we concentrate our efforts on studies of the effective potential; the problem of the possible anomalies in our model will be treated elsewhere.

In this section, we aim to calculate the exact bumblebee potential for the $B_\mu$. To do this, we start with the generating functional
\begin{eqnarray}
Z(\bar \eta,\,\eta) &=& \int DB_\mu D\psi D\bar\psi e^{i\int d^4x({\cal L}+\bar\eta\psi+\bar\psi\eta)}\nonumber\\
&=& \int DB_\mu e^{i\int d^4x \frac{g^2}{2}B_\mu B^\mu}\int D\psi D\bar\psi e^{i\int d^4x(\bar\psi S^{-1}\psi+\bar\eta\psi+\bar\psi\eta)},
\end{eqnarray}
where $S^{-1}=i\slashed{\partial}-m-ie(\partial_\mu B^\mu)\gamma_5-2ieB^\mu\partial_\mu\gamma_5$ is the operator characterizing the quadratic action, i.e., the inverse of the propagator. Next, we perform a shift of the fermionic fields, $\psi\rightarrow \psi-S\eta$ and $\bar{\psi}\rightarrow \bar{\psi}-\bar{\eta}S$, which leads to $\bar\psi S^{-1}\psi+\bar\eta\psi+\bar\psi\eta \rightarrow \bar\psi S^{-1}\psi-\bar\eta S \eta$. This results in
\begin{eqnarray}
Z(\bar \eta,\,\eta) &=& \int DB_{\mu} e^{i\int d^4x \frac{g^2}{2}B_\mu B^\mu}\int D\psi D\bar\psi e^{i\int d^4x(\bar\psi S^{-1}\psi-\bar\eta S \eta)}.
\end{eqnarray}
Finally, after integrating over the fermions, we obtain
\begin{equation}\label{GF}
Z(\bar \eta,\,\eta) = \int DB_\mu \exp\left(iS_\mathrm{eff}[B] - i\int d^4x\, \bar\eta\, S\, \eta \right),
\end{equation}
where the effective action is given by
\begin{equation}\label{Seff}
S_\mathrm{eff}[B] = \frac{g^2}{2} \int d^4x\,B_\mu B^\mu - i\mathrm{Tr}\ln(\slashed{p}-m-ie(\partial_\mu B^\mu)\gamma_5-2eB^\mu p_\mu\gamma_5).
\end{equation}
The symbol $\mathrm{Tr}$ represents the trace over Dirac matrices and the integration over momentum or coordinate spaces. Using the key identity of the derivative expansion method \cite{Das} $B^\mu p_\mu=(p-i\partial)_\mu B^\mu$, we can rewrite the above expression as
\begin{equation}\label{Seff1}
S_\mathrm{eff}[B] = \frac{g^2}{2} \int d^4x\,B_\mu B^\mu - i\mathrm{Tr}\ln(\slashed{p}-m-e(2p-i\partial)_\mu B^\mu\gamma_5).
\end{equation}
Thus, the effective potential is
\begin{equation}\label{Vef}
V_\mathrm{eff} = -\frac{g^2}{2}B_\mu B^\mu + i\mathrm{tr}\int\frac{d^4p}{(2\pi)^4}\,\ln(\slashed{p}-m-2ep_\mu B^\mu\gamma_5).
\end{equation}
We note that our study, in a certain sense, can be treated as a LV generalization of the study performed in \cite{Leblanc:1993gx}, for a vector sector.

Let us now investigate whether the potential (\ref{Vef}) has nontrivial minima by analyzing the gap equation 
\begin{equation}\label{DVef}
\frac{dV_\mathrm{eff}}{dB_\mu}\Big|_{B_\mu=\beta_\mu} = -\frac{g^2}{e} b^\mu - ie\Pi^\mu = 0,
\end{equation}
where $\beta_\mu=\frac{b_\mu}{e}$ and $\Pi^\mu$ is the one-loop tadpole amplitude given by 
\begin{equation}\label{Tad}
\Pi^\mu = \mathrm{tr} \int\frac{d^4p}{(2\pi)^4} \frac{1}{\slashed{p}-2p_\alpha b^\alpha\gamma_5-m} 2p^\mu\gamma_5.
\end{equation} 
Using the rationalized propagator, obtained through the rule
\begin{equation}
\frac{1}{\slashed{p}-2(p\cdot b)\gamma_5-m} = \frac{\slashed{p}-2(p\cdot b)\gamma_5+m}{p^2+4(p\cdot b)^2-m^2},
\end{equation}
and calculating the trace over Dirac matrices, we obtain 
\begin{equation}\label{Tad2}
\Pi^\mu = \int\frac{d^4p}{(2\pi)^4} \frac{-16(p\cdot b)p^\mu}{p^2+4(p\cdot b)^2-m^2}.
\end{equation}

To calculate the above expression in the perturbative approach, we need to expand the propagator in terms of $b_{\mu}$. This means that the $D$-dimensional tadpole amplitude (\ref{Tad2}) can be expressed as a power series in $b_{\mu}$, i.e., $\Pi_{\mu} = \sum\limits_n \Pi_{\mu}^{(n)}$, where
\begin{equation}
\Pi_{\mu}^{(n)} = \mu^{4-D}\int\frac{d^Dp}{(2\pi)^D} \frac{(-1)^n2^{2n+2}}{(p^2-m^2)^n}(p\cdot b)^{2n-1}p_\mu,
\end{equation}
 with only odd $n$ yielding nontrivial contributions. Then, for $n=1$ and $n=3$, we have
\begin{eqnarray}
-ie\Pi_\mu^{(1)} = -\frac{e m^4}{2\pi^2} \left(\frac{1}{\epsilon'}+\frac{3}{4} \right) b_\mu
\end{eqnarray}
and
\begin{eqnarray}
-ie\Pi_\mu^{(3)} = \frac{3e m^4}{\pi^2} \left(\frac{1}{\epsilon'}+\frac{3}{4} \right) b^2 b_\mu,
\end{eqnarray}
respectively, where $\frac 1{\epsilon'}=\frac 1\epsilon-\ln\frac m{\mu'}$, with $\epsilon=4-D$ and $\mu'^2=4\pi\mu^2e^{-\gamma}$. Therefore, the gap equation (\ref{DVef}) takes the form
\begin{equation}\label{DVef2}
\frac{dV_\mathrm{eff}}{dB_\mu}\Big|_{B_\mu=\beta_\mu} = \left[-\frac{e}{G} -\frac{em^4}{2\pi^2}\left(\frac{1}{\epsilon'}+\frac{3}{4} \right)(1-6b^2)\right]b^\mu +\cdots = 0.
\end{equation}

Our next step is to calculate the tadpole amplitude (\ref{Tad2}) using the nonperturbative approach. To do this, we rewrite Eq.~(\ref{Tad2}) as
\begin{equation}\label{Tad3}
\Pi^\mu = -16\int\frac{d^4p}{(2\pi)^4} \frac{p^\mu p^\nu b_\mu}{p'^2-m^2},
\end{equation}
where $p'_\alpha=M_{\alpha\beta}p^\beta$, with $M_{\alpha\beta}=g_{\alpha\beta}+\beta b_\alpha b_\beta$ and 
\begin{equation}\label{beta}
\beta = \frac{-1+\sqrt{1+4b^2}}{b^2}.
\end{equation}
After that, we have $d^4p'=\mathrm{det}\left(\frac{\partial p'^\mu}{\partial p^\nu}\right)d^4p$, i.e., $d^4p'=\mathrm{det}\left(M^{\mu\alpha}g_{\alpha\nu}\right)d^4p=-\mathrm{det}\left(M^{\mu\alpha}\right)d^4p$, so that 
\begin{equation}
d^4p = -\mathrm{det}^{-1}\left(M^{\mu\alpha}\right) d^4p'.
\end{equation}
We note that this change of variables will not affect the effective action since the functional integral over the fermions has already been performed, and for the vector field, it will yield only a linear transformation, and therefore, only a trivial modification of the generating functional, implying, at most, in arising an additive constant.
Since $p_\alpha=(M^{-1})_{\alpha\beta}p'^\beta$, we must calculate $(M^{-1})_{\alpha\beta}$, which yields the result
\begin{equation}\label{Mm1}
(M^{-1})_{\alpha\beta} = g_{\alpha\beta}-\frac{\beta}{1+\beta b^2}b_\alpha b_\beta. 
\end{equation}
Therefore, as it can be easily found that $\mathrm{det}(M^{\kappa\lambda}) = -(1+\beta b^2)$, Eq.~(\ref{Tad2}) becomes
\begin{equation}\label{Tad4}
\Pi^{\mu} = -16(1+\beta b^2)^{-1}(M^{-1})^{\mu\alpha}(M^{-1})^{\nu\beta}b_\nu \int\frac{d^4p'}{(2\pi)^4} \frac{p'_\alpha p'_\beta}{p'^2-m^2}.
\end{equation}
The solution of the above integral in $D$ dimensions is
\begin{equation}\label{intD}
\mu^{4-D}\int\frac{d^Dp'}{(2\pi)^D} \frac{p'_\alpha p'_\beta}{p'^2-m^2} = \frac{im^4}{32\pi^2}\left(\frac{1}{\epsilon'}+\frac{3}{4}\right)g_{\alpha\beta},
\end{equation}
so that, after the contraction
\begin{equation}\label{ct}
(1+\beta b^2)^{-1}(M^{-1})^{\mu\alpha}(M^{-1})^{\nu\beta}b_\nu g_{\alpha\beta} = \left(1+4b^2\right)^{-3/2}b^\mu,	
\end{equation}
we have
\begin{equation}
\Pi^{\mu} =-\frac{im^4}{2\pi^2}\left(\frac{1}{\epsilon'}+\frac{3}{4}\right)\left(1+4b^2\right)^{-3/2}b^\mu.
\end{equation}
Thus, we obtain the gap equation
\begin{eqnarray}\label{DVef3}
\frac{dV_\mathrm{eff}}{dB_\mu}\Big|_{B_\mu=\beta_\mu} = \left[-\frac{e}{G} -\frac{em^4}{2\pi^2}\left(\frac{1}{\epsilon'}+\frac{3}{4}\right)\left(1+4b^2\right)^{-3/2}\right]b^\mu = 0.
\end{eqnarray}
It is worth noting that Eq.~(\ref{DVef3}), up to the first order in $b^2$, reproduces exactly Eq.~(\ref{DVef2}), as expected. 

To simplify the solving of Eq.~(\ref{DVef3}), we introduce the renormalized mass $m_R=Z_m^{-1/4}m$, where 
\begin{equation}
\frac{1}{Z_{m}} = \frac{1}{2\pi^2} \left(\frac{1}{\epsilon'}+\frac{3}{4} \right).
\end{equation}
Hence, we can see that, for a nontrivial solution of Eq.~(\ref{DVef3}), we have 
\begin{equation}\label{1oG}
\frac{1}{G} = -m_R^4\left(1+4b^2\right)^{-3/2},
\end{equation}
which indicates that $G<0$, meaning $G=-|G|$. By solving the above expression for $b_\mu$, we get 
\begin{equation}
b^2 =\frac{1}{4}[(|G|m_R^4)^{2/3}-1].
\end{equation}
This allows us to determine that if $b_\mu$ is timelike (spacelike), $|G|>\frac{1}{m_R^4}$ ($|G|<\frac{1}{m_R^4}$).

Finally, after integrating the gap equation (\ref{DVef3}), we can find the effective potential
\begin{eqnarray}\label{Veff0}
V_\mathrm{eff} = -\frac{e^2}{2G}B_\mu B^\mu +\frac{m_R^4}{4}\left(1+4e^2B_\mu B^\mu\right)^{-1/2} -\frac{m_R^4}{4}.
\end{eqnarray}
Similarly, by substituting Eq.~(\ref{1oG}), we obtain
\begin{eqnarray}\label{Veff}
V_\mathrm{eff} =  \frac{m_R^4}{2}e^2B_\mu B^\mu\left(1+4b^2\right)^{-3/2}+\frac{m_R^4}{4}\left(1+4e^2B_\mu B^\mu\right)^{-1/2} -\frac{m_R^4}{4}.
\end{eqnarray}
Therefore, this is the exact bumblebee potential for the vector field $B_\mu$ that we are seeking, with the lowest contribution taking the form 
\begin{equation}\label{Veff2}
V_\mathrm{eff} = \frac{3m_R^4}{2} (e^2B_\mu B^\mu-b_\mu b^\mu)^2+\cdots,
\end{equation} 
where we have added the constant term $\frac{3m_R^4}{2}b^4$. We note that, unlike \cite{Gomes:2007mq}, where only lower contributions to the one-loop effective potential were calculated, here we have found it exactly, as in \cite{Assuncao:2017tnz}, despite a more complicated form of the propagator. Similar to \cite{Assuncao:2017tnz}, our effective potential (\ref{Veff}) is positively defined, providing an infinite set of minima allowing for a spontaneous Lorentz symmetry breaking. A remarkable result is the non-polynomial form of our effective potential.

Let us now study the critical temperature at which the Lorentz symmetry is restored. Assuming from now on that the system is in thermal equilibrium with a temperature $T=\beta^{-1}$, we change the expression (\ref{intD}) from Minkowski space to Euclidean space. We split the momentum $p'_\mu$ in its  spatial and temporal components, performing the following replacements: 
\begin{equation}
g^{\mu\nu}\to-\delta^{\mu\nu},\ \mathrm{i.e.},\  p'^2\to-p'^2,\ p'_\mu \to -p'_\mu,
\end{equation}
\begin{equation}
\mu^{4-D}\int \frac{d^Dp'}{(2\pi)^D} \to \mu^{3-d}\int \frac{d^d\vec p}{(2\pi)^d} \,i\int \frac{dp_0}{2\pi},
\end{equation}
and 
\begin{equation}
p'^\mu=\vec{p}^\mu + p_0 u^\mu,
\end{equation}
where $\vec p^\mu=(0,\vec p)$ and $u^\mu=(1,0,0,0)$, with $D=d+1$. We note that the vector $u^{\mu}$ is directed along the time axis and, therefore, has nothing to do with the $B_{\mu}$ field, whose direction is completely arbitrary.

In the thermal regime, the antiperiodic boundary conditions for fermions lead to discrete values of $p_0$, i.e., $p_0 = (2n+1)\frac{\pi}{\beta}$, with $n$ being integer, so that ${\textstyle\int}\frac{dp_0}{2\pi}\rightarrow \frac{1}{\beta}{\textstyle\sum}_n$. Thus, we get
\begin{equation}\label{intD2}
\mu^{4-D}\int\frac{d^Dp'}{(2\pi)^D} \frac{p'_\alpha p'_\beta}{p'^2-m^2} = -i\mu^{3-d}\frac1\beta\sum_n\int\frac{d^d\vec{p}}{(2\pi)^d}\frac{\frac{\vec p^2}{d}(\delta_{\alpha\beta}-u_\alpha u_\beta)+p_0^2u_\alpha u_\beta}{\vec p^2+p_0^2+m^2},
\end{equation}
where we have considered $\vec p_\alpha \vec p_\beta \to \frac{\vec p^2}{d}(\bar\delta_{\alpha\beta}-u_\alpha u_\beta)$. After performing the momentum integration and carrying out the sum (see \cite{Ford:1979ds} for details), we obtain
\begin{eqnarray}\label{intD3}
\mu^{4-D}\int\frac{d^Dp'}{(2\pi)^D} \frac{p'_\alpha p'_\beta}{p'^2-m^2} &=& \frac{im_R^4}{16}g_{\alpha\beta} -\frac{4i\pi^2T^4}{3}\int_{|\xi|}^\infty dz (z^2-\xi^2)^{3/2}(1-\tanh(\pi z)) g_{\alpha\beta} \\
&& -\frac{4i\pi^2T^4}{3}\int_{|\xi|}^\infty dz (z^2-\xi^2)^{1/2}(4z^2-\xi^2)(1-\tanh(\pi z)) u_\alpha u_\beta, \nonumber
\end{eqnarray}
with $\xi=\frac{m}{2\pi T}$, where we have returned to Minkowski space. In the limit of high temperature (or also in the case of $m\ll T$), $\xi \to 0$, so that Eq.~(\ref{intD3}) becomes
\begin{eqnarray}\label{intD4}
\mu^{4-D}\int\frac{d^Dp'}{(2\pi)^D} \frac{p'_\alpha p'_\beta}{p'^2-m^2} &=& \frac{im_R^4}{16}g_{\alpha\beta} -\frac{7i\pi^2T^4}{720} g_{\alpha\beta} -\frac{7i\pi^2T^4}{180}u_\alpha u_\beta.
\end{eqnarray}

Thus, after considering the contractions  (\ref{ct}) and
\begin{equation}\label{ct2}
(1+\beta b^2)^{-1}(M^{-1})^{\mu\alpha}(M^{-1})^{\nu\beta}b_\nu u_\alpha u_\beta = \left(1+4b^2\right)^{-3/2}(1-\sqrt{1+4b^2})\frac{b_0^2}{b^2}b^\mu +\left(1+4b^2\right)^{-1}b_0 u^\mu,	
\end{equation}
the tadpole amplitude (\ref{Tad4}) is written as
\begin{eqnarray}
\Pi^{\mu} &=& \left[-im_R^4+\frac{7i\pi^2T^4}{45}+\frac{28i\pi^2T^4}{45}(1-\sqrt{1+4b^2})\frac{b_0^2}{b^2}\right]\left(1+4b^2\right)^{-3/2}b^\mu \nonumber\\
&&+\frac{28i\pi^2T^4}{45}\left(1+4b^2\right)^{-1}b_0 u^\mu.
\end{eqnarray}
Therefore, the gap equation is now 
\begin{eqnarray}
\frac{dV_\mathrm{eff}}{dB_\mu}\Big|_{B_\mu=\beta_\mu} &=& -\frac{e}{G}b^\mu -\left[em_R^4-\frac{7e\pi^2T^4}{45}-\frac{28e\pi^2T^4}{45}(1-\sqrt{1+4b^2})\frac{b_0^2}{b^2}\right]\left(1+4b^2\right)^{-3/2}b^\mu \nonumber\\
&&+\frac{28e\pi^2T^4}{45}\left(1+4b^2\right)^{-1}b_0 u^\mu = 0.
\end{eqnarray}
To find the critical temperature, we can consider the decomposition $b^\mu=\vec b^\mu+b_0u^\mu$ and analyze the above equation separately for $\vec b^\mu$ and $b_0$. 

Initially, for the purely space-like $b_{\mu}$, i.e., $b_0=0$, we have
\begin{eqnarray}
\frac{dV_\mathrm{eff}}{dB_\mu}\Big|_{B_\mu=\beta_\mu} = \left[-\frac{e}{G} -e\left(m_R^4-\frac{7\pi^2T^4}{45}\right)\left(1+4\vec b^2\right)^{-3/2}\right]\vec b^\mu = 0.
\end{eqnarray}
Then, after integrating the above expression, we find the same effective potential (\ref{Veff0}), with $m_R^4 \to m_R^4-\frac{7\pi^2T^4}{45}$. Thus, as $G=-|G|$, it is early to observe that the symmetry is restored when $m_R^4-\frac{7\pi^2T^4}{45} \le 0$, i.e., for $T^4 \ge \frac{45m_R^4}{7\pi^2}$, so that the critical temperature is $T_c=\frac{45m_R^4}{7\pi^2}$. 

At the same time, for the purely time-like $b_{\mu}$, i.e., $\vec b^\mu=0$, we have
\begin{eqnarray}
\frac{dV_\mathrm{eff}}{dB_\mu}\Big|_{B_\mu=\beta_\mu} = \left[-\frac{e}{G} -e\left(m_R^4-\frac{7\pi^2T^4}{9}\right)\left(1+4 b_0^2\right)^{-3/2}\right]\vec b_0u^\mu = 0,
\end{eqnarray}
which has the same qualitative behavior as the previous case, with the only difference being in the value of the critical temperature. In this case, we arrive at $T^4 \ge \frac{9m_R^4}{7\pi^2}$, with the critical temperature $T_c=\frac{9m_R^4}{7\pi^2}$.

\section{Effective action}\label{action}

In this section, we will study the one-loop low-energy effective action for our vector field. To do so, we rewrite the Eq.~(\ref{Seff}) as follows:
\begin{equation}\label{Seff2}
S_\mathrm{eff}[B] = \frac{g^2}{2} \int d^4x\, B_\mu B^\mu +\sum_{n=1}^{\infty} S^{(n)}_\mathrm{eff}[B],
\end{equation}
where
\begin{equation}\label{Seffn}
S^{(n)}_\mathrm{eff}[B]= \frac{i}{n}\mathrm{Tr}\left[S(p)e(2p-i\partial)_\mu B^\mu\gamma_5\right]^n
\end{equation}
and $S(p)=(\slashed{p}-m)^{-1}$. The term $-i\mathrm{Tr} \ln(\slashed{p}-m)$, which is independent of $B_\mu$, has been disregarded.

The $n=1$ (tadpole) and $n=3$ (three-point) contributions vanish, i.e., $S^{(1)}_\mathrm{eff}[B]=0$ and $S^{(3)}_\mathrm{eff}[B]=0$. Then, let us focus on contributions with $n=2$ and $n=4$, whose analysis is sufficient for generating the kinetic and the lower-order potential terms. Initially, for $n=2$, we have
\begin{equation}
S^{(2)}_\mathrm{eff}[B] = \frac i2 \mathrm{Tr}\, S(p)e(2p-i\partial)_\mu B^\mu\gamma_5S(p)e(2p-i\partial)_\nu B^\nu\gamma_5 = \frac{ie^2}{2} \int d^4x\, \Pi^{\mu\nu}B_\mu B_\nu,
\end{equation}
where
\begin{equation}
\Pi^{\mu\nu} = \mathrm{tr} \int \frac{d^4p}{(2\pi)^4} S(p)(2p-i\partial)^\mu\gamma_5S(p)(2p-i\partial)^\nu\gamma_5.
\end{equation}

After applying the Feynman parametrization and calculating the trace and the integral of the above expression, we obtain
\begin{eqnarray}
{\cal L}^{(2)}_\mathrm{eff} &=& \frac{e^2m^4}{4 \pi ^2 \epsilon'}B_\mu B^\mu-\frac{e^2}{24 \pi^2 \epsilon'}B_\mu(g^{\mu\nu}k^2-k^\mu k^\nu)(6m^2-k^2)B_\nu +\frac{3 e^2m^4}{16 \pi ^2}B_\mu B^\mu \nonumber\\
&&- \frac{e^2}{72 \pi ^2}B_\mu(g^{\mu\nu}k^2-k^\mu k^\nu)\left[21 m^2-4 k^2-3 k^2\left(\frac{4 m^2}{k^2}-1\right)^{3/2}\mathrm{csc}^{-1}\left(\frac{2m}{\sqrt{k^2}}\right)\right]B_\nu,
\end{eqnarray}
with $k_{\mu} = i\partial_{\mu}$, where we have considered $S^{(2)}_\mathrm{eff}[B]=\int d^4x {\cal L}^{(2)}_\mathrm{eff}$. Thus, in the low-energy limit ($k^2\ll m^2$), we get  
\begin{eqnarray}
{\cal L}^{(2)}_\mathrm{eff} &=& \frac{e^2m^2}{4\pi^2}\left(\frac{1}{\epsilon'}+\frac{1}{2}\right)B_\mu(g^{\mu\nu}\partial^2-\partial^\mu \partial^\nu)B_\nu+\frac{e^2m^4}{4 \pi ^2}\left(\frac{1}{\epsilon'}+\frac{3}{4}\right)B_\mu B^\mu+{\cal O}\left(\partial^4\right).
\end{eqnarray}
This expression can be rewritten as
\begin{eqnarray}
{\cal L}^{(2)}_\mathrm{eff} &=& \frac{1}{2Z_3}B_\mu(g^{\mu\nu}\partial^2-\partial^\mu \partial^\nu)B_\nu+\frac{e^2m_R^4}{2}B_\mu B^\mu+{\cal O}\left(\partial^4\right),
\end{eqnarray}
where
\begin{equation}
\frac{1}{Z_{3}} = \frac{e^2m^2}{2\pi^2} \left(\frac{1}{\epsilon'}+\frac{1}{2}\right).
\end{equation}
Defining the renormalized field $B_R^\mu=Z_3^{-1/2}B^\mu$ and the coupling constant $e_R=Z_3^{1/2}e$, we arrive at
\begin{eqnarray}\label{Seff2a}
{\cal L}^{(2)}_\mathrm{eff} &=& -\frac{1}{4} F_{R\mu\nu}F_R^{\mu\nu} +\frac{e_R^2m_R^4}{2} B_{R\mu} B_R^{\mu},
\end{eqnarray}
where we have omitted the higher-order terms.

Now, for $n=4$, we have
\begin{eqnarray}
S^{(4)}_\mathrm{eff}[B] &=& \frac{i}{4} \mathrm{Tr}\, S(p)e(2p-i\partial)_\kappa B^\kappa\gamma_5S(p)e(2p-i\partial)_\lambda B^\lambda\gamma_5S(p)e(2p-i\partial)_\mu B^\mu\gamma_5 \nonumber\\
&& \times S(p)e(2p-i\partial)_\nu B^\nu\gamma_5 = \frac{ie^4}{4} \int d^4x\, \Pi^{\kappa\lambda\mu\nu} B_\kappa B_\lambda B_\mu B_\nu,
\end{eqnarray}
where
\begin{equation}
\Pi^{\kappa\lambda\mu\nu} = \mathrm{tr} \int \frac{d^4p}{(2\pi)^4} S(p)2p^\kappa\gamma_5S(p)2p^\lambda\gamma_5S(p)2p^\mu\gamma_5S(p)2p^\nu\gamma_5 +{\cal O}\left(\partial^4\right).
\end{equation}
Then, after performing the trace and integral (and omitting the higher-order terms), the resulting fourth-order effective Lagrangian is 
\begin{equation}
{\cal L}^{(4)}_\mathrm{eff} = -\frac{3e^4m^4}{4\pi^2} \left(\frac{1}{\epsilon'}+\frac{3}{4}\right) B_\mu B^\mu B_\nu B^\nu,
\end{equation}
or, in terms of the normalized quantities,
\begin{equation}\label{Seff4}
{\cal L}^{(4)}_\mathrm{eff} = -\frac{3e_R^4m_R^4}{2} B_{R\mu} B_R^\mu B_{R\nu} B_R^\nu.
\end{equation}

Finally, the Lagrangian of (\ref{Seff2}), with the results (\ref{Seff2}) and (\ref{Seff4}), is given by
\begin{eqnarray}
{\cal L}_{B} &=& \frac{e_R^2}{2G}B_{R\mu}B_R^{\mu} -\frac{1}{4} F_{R\mu\nu}F_R^{\mu\nu} +\frac{e_R^2m_R^4}{2} B_{R\mu}B_R^{\mu} -\frac{3e_R^4m_R^4}{2} B_{R\mu}B_R^{\mu}B_{R\nu}B_R^{\nu}.
\end{eqnarray}
We can simplify this expression by expanding (\ref{1oG}) (up to the first order in $b^2$), i.e., by considering $\frac1G=-m_R^4(1-6b_{\mu}b^{\mu})$. So, we obtain
\begin{eqnarray}\label{LB}
{\cal L}_{B} &=& -\frac{1}{4} F_{R\mu\nu}F_R^{\mu\nu} -\frac{3 m_R^4}{2} (e_R^2B_{R\mu}B_R^{\mu}-b_{\mu}b^{\mu})^2,
\end{eqnarray}
where we have added the constant $-\frac{3m_R^4}{2}b^4$. Thus, we have generated the low-energy effective Lagrangian for the vector field, which is given by the usual Maxwell kinetic term and the bumblebee-like potential.

\section{Summary}\label{summary}

Starting with a new massive four-fermion model with derivative-dependent coupling (\ref{L0}), essentially distinct from that one introduced in our earlier paper \cite{Assuncao:2017tnz},  we generated the bumblebee effective Lagrangian as a one-loop contribution to the effective action. We explicitly demonstrated that our effective potential is positively definite, implying the existence of an infinite set of nontrivial minima, which results in the spontaneous breaking of Lorentz symmetry. A very interesting consequence of non-zero mass of fermions consists in non-polynomiality of our effective potential (see Eq.~(\ref{Veff})), which also represents one more difference from the results for the massless case found in \cite{Assuncao:2017tnz}. Additionally, we calculated the effective Lagrangian (\ref{LB}), which can then be expressed in terms of the renormalized quantities $m_R$, $e_R$, and $B_R^\mu$.

Moreover, we studied the finite-temperature behavior of our effective potential and demonstrated that our model displays a possibility of phase transitions for both time-like and space-like LV vectors.

The natural continuation of this study can consist in its generalization for the case of the antisymmetric tensor field, including studies of possibility to generate a stable LV vector-tensor model proposed in \cite{Potting2023}, and to study the impact of finite temperature on stability issues for this class of models. Besides,  it is very interesting to consider this theory on a curved space-time background. We expect to perform this study in forthcoming papers.

{\bf Acknowledgments.}  The work of T. M. has been partially supported by the CNPq project No. 316499/2021-8 and FAPEAL project No. E:60030.0000002341/2022. The work of A. Yu.\ P. has been partially supported by the CNPq project No. 303777/2023-0.


\begin{thebibliography}{100}

\bibitem{Bjorken:1963vg}
J.~D.~Bjorken,
Annals Phys. \textbf{24}, 174-187 (1963).

\bibitem{Bjorken:2001pe}
J.~Bjorken,
[arXiv:hep-th/0111196 [hep-th]].

\bibitem{Bialynicki-Birula:1963usy}
I.~Bialynicki-Birula,
Phys. Rev. \textbf{130}, 465-468 (1963).

\bibitem{Guralnik:1964zz}
G.~S.~Guralnik,
Phys. Rev. \textbf{136}, B1404-B1416 (1964).

\bibitem{Phillips:1966zzc}
P.~R.~Phillips,
Phys. Rev. \textbf{146}, 966-973 (1966).

\bibitem{Eguchi:1976iz}
T.~Eguchi,
Phys. Rev. D \textbf{14}, 2755 (1976).

\bibitem{Amati:1981xt}
D.~Amati, R.~Barbieri, A.~C.~Davis and G.~Veneziano,
Phys. Lett. B \textbf{102}, 408-412 (1981).

\bibitem{Andrianov:1998ay}
A.~A.~Andrianov, R.~Soldati and L.~Sorbo,
Phys. Rev. D \textbf{59}, 025002 (1999)
[arXiv:hep-th/9806220 [hep-th]].

\bibitem{Kraus:2002sa}
P.~Kraus and E.~T.~Tomboulis,
Phys. Rev. D \textbf{66}, 045015 (2002)
[arXiv:hep-th/0203221 [hep-th]].

\bibitem{Jenkins:2003hw}
A.~Jenkins,
Phys. Rev. D \textbf{69}, 105007 (2004)
[arXiv:hep-th/0311127 [hep-th]].

\bibitem{Inagaki:1997kz}
T.~Inagaki, T.~Muta and S.~D.~Odintsov,
Prog. Theor. Phys. Suppl. \textbf{127} (1997), 93
[arXiv:hep-th/9711084 [hep-th]].

\bibitem{Buchbinder:1997ta}
I.~L.~Buchbinder, T.~Inagaki and S.~D.~Odintsov,
Mod. Phys. Lett. A \textbf{12} (1997), 2271-2278
[arXiv:hep-th/9702097 [hep-th]].

\bibitem{Elizalde:1998sv}
E.~Elizalde, S.~P.~Gavrilov, S.~D.~Odintsov and Y.~I.~Shil'nov,
Braz. J. Phys. \textbf{30} (2000), 573-580
[arXiv:hep-ph/9807368 [hep-ph]].

\bibitem{Leblanc:1993gx}
M.~Leblanc, R.~MacKenzie, P.~K.~Panigrahi and R.~Ray,
Int. J. Mod. Phys. A \textbf{9} (1994), 4717-4726
[arXiv:hep-th/9311076 [hep-th]].


\bibitem{Kostelecky:1989jp}
V.~A.~Kostelecky and S.~Samuel,
Phys. Rev. Lett. \textbf{63}, 224 (1989).

\bibitem{Kostelecky:1989jw}
V.~A.~Kostelecky and S.~Samuel,
Phys. Rev. D \textbf{40}, 1886-1903 (1989).

\bibitem{ColKost1} 
D.~Colladay and V.~A.~Kostelecky,
Phys. Rev. D \textbf{55} (1997), 6760-6774
[arXiv:hep-ph/9703464 [hep-ph]].

\bibitem{ColKost2} D.~Colladay and V.~A.~Kostelecky,
Phys. Rev. D \textbf{58} (1998), 116002
[arXiv:hep-ph/9809521 [hep-ph]].

\bibitem{KosGra} V.~A.~Kostelecky,
Phys. Rev. D \textbf{69} (2004), 105009
[arXiv:hep-th/0312310 [hep-th]].

\bibitem{KosLi} V.~A.~Kosteleck\'y and Z.~Li,
Phys. Rev. D \textbf{103} (2021) no.2, 024059
[arXiv:2008.12206 [gr-qc]].

\bibitem{ourLV} A.~F.~Ferrari, J.~R.~Nascimento and A.~Y.~Petrov,
Eur. Phys. J. C \textbf{80} (2020) no.5, 459
[arXiv:1812.01702 [hep-th]].

\bibitem{Gomes:2007mq}
M.~Gomes, T.~Mariz, J.~R.~Nascimento and A.~J.~da Silva,
Phys. Rev. D \textbf{77}, 105002 (2008)
[arXiv:0709.2904 [hep-th]].

\bibitem{Schwinger} 
J.~Schwinger,
Phys. Rev. \textbf{167} (1968), 1432-1436.

\bibitem{Kostelecky:2000mm}
V.~A.~Kostelecky and R.~Lehnert,
Phys. Rev. D \textbf{63}, 065008 (2001)
[arXiv:hep-th/0012060 [hep-th]].

\bibitem{Altschul:2005mu}
B.~Altschul and V.~A.~Kostelecky,
Phys. Lett. B \textbf{628}, 106-112 (2005)
[arXiv:hep-th/0509068 [hep-th]].

\bibitem{tHooft:1972tcz}
G.~'t Hooft and M.~J.~G.~Veltman,
Nucl. Phys. B \textbf{44}, 189-213 (1972).

\bibitem{Assuncao:2017tnz}
J.~F.~Assun\c{c}ao, T.~Mariz, J.~R.~Nascimento and A.~Y.~Petrov,
Phys. Rev. D \textbf{96}, no.6, 065021 (2017)
[arXiv:1707.07778 [hep-th]].

\bibitem{Adler:1969gk}
S.~L.~Adler,
Phys. Rev. \textbf{177} (1969), 2426-2438

\bibitem{Bell:1969ts}
J.~S.~Bell and R.~Jackiw,
Nuovo Cim. A \textbf{60} (1969), 47-61

\bibitem{Assuncao:2019azw}
J.~F.~Assun\c{c}\~ao, T.~Mariz, J.~R.~Nascimento and A.~Y.~Petrov,
Phys. Rev. D \textbf{100}, no.8, 085009 (2019)
[arXiv:1902.10592 [hep-th]].

\bibitem{Das} 
K.~S.~Babu, A.~K.~Das and P.~Panigrahi,
Phys. Rev. D \textbf{36}, 3725 (1987).

\bibitem{Ford:1979ds} 
  L.~H.~Ford,
  Phys.\ Rev.\ D {\bf 21}, 933 (1980).
  
 \bibitem{Potting2023} R.~Potting,
[arXiv:2311.03159 [hep-th]].

\end{thebibliography}
\end{document}